\documentclass{elsart}

\usepackage{graphicx}
\begin{document}

\begin{frontmatter}


\title{THE INTEGRAL EQUATION APPROACH TO KINEMATIC DYNAMO THEORY 
AND ITS APPLICATION TO DYNAMO EXPERIMENTS IN 
CYLINDRICAL GEOMETRY}

\author{M. Xu\thanksref{now}},
\thanks[now]{Present address: Center for Space Thermal Science, Shandong University, P.O.Box 88, Jin Shi Road
73, Jinan City, Shandong Province, P. R. China}
\author{F. Stefani\corauthref{cor1}},
\corauth[cor1]{Corresponding author. Tel.: +49 351 260 3069: fax:  +49 351 260 2007.}
\ead{F.Stefani@fzd.de}
\author{G. Gerbeth}

\address{Forschungszentrum Dresden-Rossendorf, P.O. Box 510119,
D-01314 Dresden, Germany}

\begin{abstract}
The conventional magnetic induction equation that governs hydromagnetic dynamo 
action is transformed into an equivalent integral equation system. An advantage of 
this approach is that the 
computational domain is restricted to the region occupied by the electrically 
conducting fluid and to its boundary. This integral equation 
approach is first employed to simulate kinematic dynamos excited by
Beltrami-like flows in a finite cylinder. The impact of externally
added layers around the cylinder on the onset of dynamo actions is
investigated. Then it is applied to simulate dynamo experiments within
cylindrical geometry including the ''von K\'{a}rm\'{a}n sodium'' (VKS) experiment
and the Riga dynamo experiment. A modified version of this approach is
utilized to investigate magnetic induction effects under the influence of 
externally applied magnetic fields which is also important to measure 
the proximity of a given dynamo facility to the self-excitation threshold.
\end{abstract}
\end{frontmatter}
\section{INTRODUCTION}
Dynamo action in moving electrically conducting fluids 
explains the existence of cosmic magnetic fields,
including the fields of planets, stars, and galaxies \cite{HORU}. 
As long as the magnetic field is weak and its influence on the velocity
field is negligible
we speak about the  {\it kinematic dynamo regime}. When the magnetic field has 
gained higher amplitudes the velocity field will be modified, and the dynamo enters 
its saturation regime. 

The usual way to simulate dynamos numerically is
based on the induction equation 
for the magnetic field $\textbf{B}$,
\begin{eqnarray}{\label{in1}}
\frac{\partial \textbf{B}}{\partial t}=\nabla
\times(\textbf{u}\times\textbf{B})+\frac{1}{\mu\sigma}\Delta \textbf{B},\;\;\nabla\cdot \textbf{B}=0,
\end{eqnarray}
where $\textbf{u}$ is the given velocity field, $\mu$ the permeability of the fluid, 
and $\sigma$ its electrical 
conductivity. The behaviour of the magnetic field $\textbf{B}$ in Eq. (\ref{in1}) is 
controlled by the ratio of field production and field dissipation,
expressed by the
magnetic Reynolds number $R_m=\mu\sigma LU$, where
$L$ and $U$ are typical length and velocity scales of the flow, respectively. 
When the magnetic Reynolds number reaches a critical value,
henceforth denoted by $R_m^c$, the dynamo starts to operate.

Equation (\ref{in1}) follows directly from
pre-Maxwell's equations and Ohm's law in
moving conductors. 
In order to make this equation solvable, boundary conditions of the magnetic field must be 
prescribed. In the case of vanishing excitations of the magnetic field from outside the 
considered finite region, the boundary condition of the magnetic field is given as follows:
\begin{eqnarray}{\label{in2}}
\textbf{B}=O(r^{-3})\; \mbox{as}\; r\rightarrow\infty.
\end{eqnarray}
Kinematic dynamos are usually simulated in the framework of the differential 
equation approach by solving the induction equation (\ref{in1}). For spherical 
dynamos, as they occur in planets and stars, the problem of implementing the 
non-local boundary conditions for the magnetic field is easily solved by 
using decoupled boundary conditions for each degree of the spherical harmonics. 
For other than spherically shaped dynamos, in particular for galactic
dynamos and
some of the recent laboratory dynamos working in cylindrical geometry \cite{RMP}, 
the handling of the non-local boundary conditions is a notorious problem. 

The simplest way to circumvent this problem is to replace the non-local boundary 
conditions by simplified local ones (so-called
vertical field condition). This is often
used in the simulation of 
galactic dynamos \cite{Brand}, and has also been tested in an approximate simulation 
of the Riga dynamo experiment \cite{Kenjeres}. 

For the simulation of the cylindrical Karlsruhe dynamo experiment, the actual electrically 
conducting region was embedded into a sphere, and the region between the sphere and 
the surface of the dynamo was virtually filled by a medium of lower electrical 
conductivity \cite{Radl1,Radl2}. 

Of course, both methods are connected with losses of accuracy. In order to fully implement 
the nonlocal boundary condition, Maxwell's equations must be fulfilled in the exterior, 
too. This can be implemented in different ways.
For the finite difference simulation of the Riga dynamo, the 
Laplace equation was solved (for each time-step) in the
exteriour of the dynamo domain and the magnetic field solutions 
in the interiour and in the exterior were matched using interface 
conditions \cite{Stef1}. A similar method, although based on the finite 
element method, was presented by Guermond et al. \cite{GUERMOND,LAGUERRE}. 
Another, and  quite elegant, technique to circumvent the solution in 
the exteriour was presented  
by Iskakov et al. \cite{ISKAKOV1,ISKAKOV2}  
where a  combination of a finite volume and a boundary 
element method was used to circumvent the discretization of the outer domain.

An  alternative to the differential equation approach (DEA) based on the
solution of the induction equation is the integral
equation approach (IEA) for kinematic dynamos
which basically relies on the self-consistent treatment of Biot-Savart's law. 
For the case of a steady dynamo acting in  infinite domains of homogeneous 
conductivity, the integral equation approach had already been employed by 
a few  authors \cite{Gail1,Gail2,Ya,Dobl}. For the case of finite domains, the simple 
Biot-Savart equation 
has to be supplemented by a boundary integral equation for the electric potential 
\cite{Stef2,M1}. If the 
magnetic field becomes time-dependent, yet 
another equation for the magnetic vector potential has to be added \cite{M2}. 

In the present work, the integral equation approach is applied to 
various dynamo problems in cylindrical geometry.  Two variants of the approach
are presented: in the first one,
it is implemented as an eigenvalue solver to solve genuine dynamo 
problems. In the second one, 
it is used to treat induction effects in the case of externally applied magnetic fields.
Actually, the first variant was already at the root of
the paper \cite{Stef3}
where a surprising negative 
impact of sodium layers behind the propellers in the ''von K\'{a}rm\'{a}n sodium'' (VKS) 
experiment was identified. It was not least this finding that
prompted the VKS
team to modify the experiment which made it ultimately succesful  \cite{Monc,Berh}.
After the derivation of the equation system in cylindrical geometry,
we switch over to the treatment of specific problems, 
including the free decay case, the mentioned 
''von K\'{a}rm\'{a}n sodium'' (VKS) experiment \cite{VKS1,VKS2}, and the Riga 
dynamo experiment \cite{PRL1,PRL2,PLASMA}. 

\section{MATHEMATICAL FORMULATION}
Assume the electrically conducting fluid be confined in a finite region $V$ 
with boundary $S$, the exterior of this region filled by  insulating material or vacuum. 
Then, dynamo and induction processes can be described [16]
by the following integral equation system:
\begin{eqnarray}
{\mathbf{b}}({\mathbf{r}})&=&\frac{\mu\sigma}{4\pi}\int_V\frac{({\mathbf{u}}({\mathbf{r}}')
\times({\mathbf{B}}_0({\mathbf{r}}')+{\mathbf{b}}({\mathbf{r}}')))\times({\mathbf{r}}-{\mathbf{r}}')}
{|{\mathbf{r}}-{\mathbf{r}}'|^3}dV' \nonumber\\
&&-\frac{\mu\sigma\lambda}{4\pi}\int_V\frac{{\mathbf{A}}({\mathbf{r}}')
\times({\mathbf{r}}-{\mathbf{r}}')}{|{\mathbf{r}}-{\mathbf{r}}'|^3}
dV'-\frac{\mu\sigma}{4\pi}\int_S\phi({\mathbf{s}}'){\mathbf{n}}({\mathbf{s}}')\times
\frac{{\mathbf{r}}-{\mathbf{s}}'}{|{\mathbf{r}}-{\mathbf{s}}'|^3}dS'\label{eqa1}\\
\frac{1}{2}\phi({\mathbf{s}})&=&\frac{1}{4\pi}\int_V\frac{({\mathbf{u}}({\mathbf{r}}')
\times({\mathbf{B}}_0({\mathbf{r}}')+{\mathbf{b}}({\mathbf{r}}')))\cdot({\mathbf{s}}-{\mathbf{r}}')}
{|{\mathbf{s}}-{\mathbf{r}}'|^3}dV'\nonumber\\
&&-\frac{\lambda}{4\pi}\int_V\frac{{\mathbf{A}}({\mathbf{r}}')
\cdot({\mathbf{s}}-{\mathbf{r}}')}{|{\mathbf{s}}-{\mathbf{r}}'|^3}dV'-
\frac{1}{4\pi}\int_S\phi({\mathbf{s}}'){\mathbf{n}}({\mathbf{s}}')\cdot\frac{{\mathbf{s}}-{\mathbf{s}}'}{|{\mathbf{s}}-{\mathbf{s}}'|^3}dS'\label{eqa2}\\
{\mathbf{A}}({\mathbf{r}})&=&\frac{1}{4\pi}\int_V\frac{({\mathbf{B}}_0({\mathbf{r}}')+{\mathbf{b}}({\mathbf{r}}'))\times({\mathbf{r}}-{\mathbf{r}}')}{|{\mathbf{r}}-{\mathbf{r}}'|^3}dV'\nonumber\\
&&+\frac{1}{4\pi}\int_S{\mathbf{n}}({\mathbf{s}}')\times\frac{{\mathbf{B}}_0({\mathbf{s}}')+{\mathbf{b}}({\mathbf{s}}')}{|{\mathbf{r}}-{\mathbf{s}}'|}dS',\label{eqa3}
\end{eqnarray}
where ${\mathbf{B}_0}$ is the externally applied magnetic field (which might be zero), 
${\mathbf{b}}$  the induced magnetic field, 
${\mathbf{u}}$ the 
velocity field, $\mu$ the permeability of the fluid (which is
in most relevant cases the permeability of the vacuum),
$\sigma$ the electrical conductivity, ${\mathbf{A}}$ the vector potential,
and $\phi$ the electric potential. ${\mathbf{n}}$ denotes the outward directed unit 
vector at the boundary $S$. For a steady velocity field, the time
dependence of
all electromagnetic fields can be assumed to be
$\sim \exp{\lambda t}$. We have to distinguish three different cases:
For non-zero $\mathbf{B}_0$, and below
the self-excitation threshold, the imaginary part of $\lambda$ is simply
the angular frequency of the applied and also of the induced
magnetic field. For $\mathbf{B}_0=0$ the equation system (3-5) represents
an eigenvalue equation for the unknown time constant $\lambda$
whose real part is the
growth rate, and its imaginary part the angular
frequency of the fields.
For  $\mathbf{B}_0=0$ and $\lambda=0$, we need only
the equations (3) and (4) which then
represent an eigenvalue problem for the
critical value of the velocity $\mathbf u$ at
which the (non-oscillatory) dynamo
starts to work.

\subsection{General numerical scheme}
Although, in this paper, we will focus mainly on cylindrical
systems it might be instructive to delineate the
general numerical scheme for the solution of Eqs. (3-5).

Assuming a specific discretization
of all fields in Eqs. (\ref{eqa1}-\ref{eqa3}), we obtain
\begin{eqnarray}
b_i&=&\mu\sigma[P_{ik}(B_{0k}+b_k)-\lambda R_{ij}A_j-Q_{il}\phi_l],\label{eqg1}\\
G_{ml}\phi_l&=&S_{mk}(B_{0k}+b_k)-\lambda T_{mj}A_j,\label{eqg2}\\
A_j&=&W_{jk}(B_{0k}+b_k),\label{eqg3}
\end{eqnarray}
where Einstein's summation convention is 
assumed. We have used the notion
$G_{ml}=0.5  \, \delta_{ml}+U_{ml}$. $B_{0k}$ and $b_k$
denote the degrees of freedom of the externally 
added magnetic field and the induced magnetic field, $A_j$ 
the degrees of freedom of the vector potential in the volume
$V$, $\phi_l$ the degrees of freedom of the electric potential 
at the boundary surface. Note that only the matrices 
$P_{ik}$ and $S_{mk}$ depend on the velocity (the sources of
the dynamo action),
while $R_{ij}$, $Q_{il}$, $T_{mj}$, $G_{ml}$ and $W_{jk}$ 
depend only on the geometry of the dynamo domain and the 
discretization details.

Substituting Eqs.(\ref{eqg2}) and (\ref{eqg3}) into Eq.(\ref{eqg1}) 
and eliminating $A_j$ and $\phi_l$ gives one single matrix equation
for the induced magnetic field components $b_i$:
\begin{eqnarray}{\label{eqg4}}
b_i&=&\mu\sigma[P_{ik}(B_{0k}+b_k)-\lambda 
R_{ij} W_{jk}(B_{0k}+b_k)-Q_{il}G_{lm}^{-1}S_{mk}(B_{0k}+b_k)\nonumber\\
&&+\lambda Q_{il}G_{lm}^{-1}T_{mj}W_{jk}(B_{0k}+b_k)].
\end{eqnarray}
This equation can be further rewritten in the following form:
\begin{eqnarray}{\label{eqg5}}
[\delta_{ik}-\mu\sigma E_{ik}-\mu\sigma\lambda F_{ik}]b_k=[\mu\sigma E_{ik}+
\mu\sigma\lambda F_{ik}]B_{0k},
\end{eqnarray}
where $E_{ik}=P_{ik}-Q_{il}G_{im}^{-1}S_{mk}$ and
$F_{ik}=-R_{ij}W_{jk}+Q_{il}G_{lm}^{-1}T_{mj}W_{jk}$.
To compute  induction effects, the induced magnetic field is 
obtained by solving the algebraic equation system (\ref{eqg5}). 
For the kinematic dynamo, Eq.(\ref{eqg5}) reduces to the 
following generalized eigenvalue problem  
\begin{eqnarray}{\label{eqg6}}
[\delta_{ik}+\mu \sigma E_{ik}]b_k=\lambda^* F_{ik}b_k,
\end{eqnarray}
where  $\lambda^*$ is a new
time constant rescaled according to
$\lambda^*=\mu\sigma \lambda$.

\subsection{Cylindrical Geometry}

Since a number of dynamo experiments are carried out in cylindrical vessels,
it is worth to specify the integral equation approach to this geometry.
As long as the dynamo source (i.e. the velocity field
or a corresponding mean-field
quantity) is axisymmetric, the different azimuthal modes
of the electromagnetic fields can be decoupled. This leads to a tremendous
reduction of the numerical effort. The price we have to pay for this
is the necessity to carefully deriving the dimensionally reduced
version of the integral equation system.

The electrically
conducting fluid is assumed to be confined in a cylinder
with radius $R$ and
height $2H$. 
Introducing the cylindrical coordinate system ($\rho, \varphi, z$), we have
\begin{eqnarray}
{\mathbf{r}}=[\rho \cos \varphi, \rho \sin \varphi, z]^T,
{\mathbf{b}}=[b_\rho, b_\varphi, b_z]^T,
{\mathbf{u}}=[u_\rho, u_\varphi, u_z]^T.\label{eq4}
\end{eqnarray}
The magnetic field ${\mathbf{b}}$, the electric potential $\phi$, and 
the vector potential ${\mathbf{A}}$ are expanded into azimuthal modes:
\begin{eqnarray}{\label{eq4a}}
\pmatrix{{\mathbf{b}}\cr
\phi\cr
{\mathbf{A}}}=\sum_{m=-\infty}^{\infty} \pmatrix{{\mathbf{b}}_m\cr \phi_m\cr {\mathbf{A}}_m} \exp(im\varphi).
\end{eqnarray}
When the velocity field is axisymmetric
(i.e. it has only a component with
$m=0$), one can see that $[{\mathbf{b}}_m, \phi_m,{\mathbf{A}}_m]^T
(m=0, \pm 1, \pm 2,\cdots)$ are decoupled with
respect to $m$ and they only depend on the variables $(\rho,z)$.
Henceforth,
we always re-denote $[{\mathbf{b}}_m, \phi_m, {\mathbf{A}}_m]^T$  
as $[{\mathbf{b}}, \phi, {\mathbf{A}}]^T$ for abbreviation. Then, after integrating over
$\varphi$, Eq.(\ref{eqa1}) acquires the form
\begin{eqnarray}
b_\rho&=&\frac{\mu\sigma}{4\pi}[\int_{-H}^H\int_{0}^{R}[((z-z')E_c^m u_z-i\rho' E_s^m u_\varphi)(B_{0\rho}+b_{\rho})+(-i(z-z')u_z E_s^m \nonumber\\
&&+i\rho' E_s^mu_\rho)(B_{0\varphi}+b_\varphi)+(i(z-z')E_s^m u_\varphi-(z-z')u_\rho E_c^m)(B_{0z}+b_z)]\rho'd\rho'dz'\nonumber\\
&&-\int_0^{R}\phi\rho'^2E_s^m|_{z'=H}d\rho'-\int_{-H}^H\phi R(z-z')E_s^m|_{\rho'=R}dz'+\int_0^{R}\phi\rho'^2E_s^m|_{z'=-H}d\rho'\nonumber\\
&&-\lambda \int_{-H}^H\int_{0}^{R}((z-z')E_s^m A_\rho+(z-z') E_c^m A_\varphi+\rho' E_s^m A_z)\rho'd\rho'dz'],\label{eq5}\\
b_\varphi&=&\frac{\mu\sigma}{4\pi}[\int_{-H}^H\int_{0}^{R}[(-(\rho E_1^m-\rho' E_c^m)u_\varphi+i(z-z')u_zE_s^m)(B_{0\rho}+b_\rho)\nonumber\\
&&+((\rho E_1^m-\rho' E_c^m)u_\rho+(z-z')u_zE_c^m)(B_{0\varphi}+b_\varphi)+(-(z-z')u_\varphi E_c^m\nonumber\\
&&-i(z-z')u_\rho E_s^m)(B_{0z}+b_z)]\rho'd\rho' dz'-\int_0^{R}\phi( \rho\rho'E_1^m|_{z'=H}-\rho'^2E_c^m|_{z'=H})d\rho'\nonumber\\
&&+\int_{-H}^H\phi R(z-z')E_c^m|_{\rho'=R}dz'-\int_0^{R}\phi(-\rho E_1^m|_{z'=-H}+\rho' E_c^m|_{z'=-H})\rho'd\rho'\nonumber\\
&&-\lambda\int_{-H}^H\int_{0}^{R}(  (z-z')E_s^m A_\varphi   -(z-z') E_c^m A_\rho +    (\rho E_1^m-\rho' E_c^m) A_z)\rho' d\rho' dz'],\label{eq6}\\
b_z&=&\frac{\mu\sigma}{4\pi}[\int_{-H}^H\int_{0}^{R} [(\rho' E_1^m-\rho E_c^m)u_z (B_{0\rho}+b_\rho) +i \rho E_s^m u_z (B_{0\varphi}+b_\varphi)\nonumber\\
&&+((-\rho' E_1^m+\rho E_c^m)u_\rho-i\rho E_s^m u_\varphi )(B_{0z}+b_z)]\rho'd\rho'dz'+\int_{-H}^H \phi R\rho E_s^m|_{\rho'=R}dz'\nonumber\\
&&-\lambda \int_{-H}^H\int_{0}^{R}(-\rho E_s^m A_\rho+(\rho' E_1^m-\rho E_c^m)A_\varphi)\rho' d\rho' dz'],\label{eq7}
\end{eqnarray}
where the following azimuthal integrals appear:
\begin{eqnarray}
E_1^m(\rho,\rho',z,z')&=&\int_0^{2\pi}\frac{\cos  m\varphi'}{(\rho^2+\rho'^2-2\rho\rho'\cos  \varphi'+(z-z')^2)^{\frac{3}{2}}}d\varphi',\nonumber\\
E_c^m(\rho,\rho',z,z')&=&\int_0^{2\pi}\frac{\cos  m\varphi'\cos  \varphi'}{(\rho^2+\rho'^2-2\rho\rho'\cos  \varphi'+(z-z')^2)^{\frac{3}{2}}}d\varphi',\nonumber\\
E_s^m(\rho,\rho',z,z')&=&\int_0^{2\pi}\frac{\sin  m\varphi'\sin  \varphi'}{(\rho^2+\rho'^2-2\rho\rho'\cos  \varphi'+(z-z')^2)^{\frac{3}{2}}}d\varphi'.\nonumber
\end{eqnarray}
Accordingly, from Eq.(\ref{eqa2}), we obtain the
expressions for the electric potentials at the three different
surface parts of the cylinder:
\begin{eqnarray}
\frac{1}{2}\phi(s_1)&=&\frac{1}{4\pi}[\int_{-H}^H\int_{0}^{R} (-\rho'\rho E_s^m|_{z=H}u_z-\rho'(H-z')u_\varphi E_1^m|_{z=H})(B_{0\rho}+b_\rho) \nonumber\\
&&+((-\rho'\rho E_c^m|_{z=H}+\rho'^2E_1^m|_{z=H})u_z+\rho'(H-z')u_\rho E_1^m|_{z=H})(B_{0\varphi}+b_\varphi) \nonumber\\
&&+((\rho'\rho E_c^m|_{z=H}-\rho'^2 E_1^m|_{z=H})u_\varphi+\rho'\rho E_s^m|_{z=H}u_\rho)(B_{0z}+b_\varphi)d\rho'dz'\nonumber\\
&&-\int_{-H}^H\phi R(\rho E_c^m|_{\rho'=R, z=H}-RE_1^m|_{\rho'=R, z=H})dz'\nonumber\\
&&+2.0H\int_0^{R}\phi E_1^m|_{z=H, z'=-H}\rho' d\rho'-\lambda \int_{-H}^H\int_{0}^{R}\rho'(\rho E_c^m|_{z=H}-\rho'E_1^m|_{z=H})A_\rho\nonumber\\
&&-\rho'\rho E_s^m|_{z=H}A_\varphi+\rho'(H-z')E_1^m|_{z=H} A_z d\rho' dz'],\label{eq8}\\
\frac{1}{2}\phi(s_2)&=&\frac{1}{4\pi}[\int_{-H}^H\int_{0}^{R}(-\rho' R E_s^m|_{\rho=R}u_z-\rho'(z-z')u_\varphi E_1^m|_{\rho=R})(B_{0\rho}+b_\rho)\nonumber\\
&&+(-\rho' R E_c^m|_{\rho=R}u_z+
\rho'^2 E_1^m|_{\rho=R}u_z+\rho'(z-z')u_\rho E_1^m|_{\rho=R})(B_{0\varphi}+b_\varphi)\nonumber\\
&&+(\rho' u_\varphi RE_c^m|_{\rho=R}+\rho' R u_\rho E_s^m|_{\rho=R}-\rho'^2 u_\varphi E_1^m|_{\rho=R})(B_{0z}+b_z)d\rho'dz'\nonumber\\
&&-\int_0^{R}\phi(z-H)E_1^m|_{\rho=R, z'=H}\rho' d\rho'-\int_{-H}^H\phi(E_c^m|_{\rho=\rho'=R}-E_1^m|_{\rho=\rho'=R})R^2
dz'\nonumber\\
&&+\nonumber\int_0^{R}\phi(z+H)E_1^m|_{\rho=R, z'=-H}\rho'd\rho'-\lambda \int_{-H}^H\int_{0}^{R}(\rho' RE_c^m|_{\rho=R}\nonumber\\
&&-\rho'^2 E_1^m|_{\rho=R}) A_\rho-
\rho' R E_s^m|_{\rho=R} A_\varphi+\rho'(z-z') E_1^m|_{\rho=R} A_zd\rho'dz'],\label{eq9}\\
\frac{1}{2}\phi(s_3)&=&\frac{1}{4\pi}[\int_{-H}^H\int_{0}^{R}(-\rho'\rho E_s^m|_{z=-H} u_z+\rho'(H+z') E_1^m|_{z=-H}u_\varphi) (B_{0\rho}+b_\rho) \nonumber\\
&&+(-\rho'\rho E_c^m|_{z=-H}u_z+\rho'^2 E_1^m|_{z=-H}u_z-\rho'(H+z')u_\rho E_1^m|_{z=-H})(B_{0\varphi}+b_\varphi)\nonumber\\
&&+(\rho'\rho u_\varphi E_c^m|_{z=-H}-\rho'^2 u_\varphi E_1^m|_{z=-H}+\rho'\rho u_\rho E_s^m|_{z=-H})(B_{0z}+b_z) d\rho'dz'\nonumber\\
&&+2.0 H\int_0^{R}\phi E_1^m|_{z=-H,z'=H}\rho' d\rho'-\int_{-H}^H \phi R(\rho E_c^m|_{\rho'=R,z=-H}\nonumber\\
&&-R E_1^m|_{\rho'=R, z=-H})dz'-\lambda\int_V(\rho\rho' E_c^m|_{z=-H}-\rho'^2 E_1^m|_{z=-H})A_\rho\nonumber\\
&&-\rho'\rho E_s^m|_{z=-H} A_\varphi+\rho'(-H-z') A_z E_1^m|_{z=-H}d\rho'dz'].\label{eq10}
\end{eqnarray}
Here, $s_1$ is the surface $z=H$, $s_2$ the surface $\rho=R$,
$s_3$ the surface $z=-H$.
Equation (\ref{eqa3}) for the vector potential gets the form
\begin{eqnarray}
A_\rho&=&\frac{1}{4\pi}[\int_{-H}^H\int_{0}^{R}\rho'(z-z')E_s^m (B_{0\rho}+b_\rho)+\rho'(z-z') E_c^m (B_{0\varphi}+b_\varphi) \nonumber\\
&&+\rho'^2E_s^m(B_{0z}+b_z)d\rho'dz'+\int_0^{R}-\rho'D_s^m|_{z'=H}(B_{0\rho}+b_\rho)-\rho'D_c^m|_{z'=H}(B_{0\varphi}+b_\varphi) d\rho'\nonumber\\
&&+\int_{-H}^HR (B_{0z}+b_z)D_s^m|_{\rho'=R} dz'+\int_{0}^{R}\rho'D_s^m|_{z'=-H}(B_{0\rho}+b_\rho)\nonumber\\ &&+\rho'(B_{0\varphi}+b_\varphi)D_c^m|_{z'=-H}d\rho']\label{eq11}\\
A_\varphi&=&\frac{1}{4\pi}[\int_{-H}^H\int_{0}^{R}-\rho'(z-z')E_c^m (B_{0\rho}+b_\rho)+\rho'(z-z') E_s^m (B_{0\varphi}+b_\varphi)\nonumber\\
&&+\rho'(\rho E_1^m-\rho' E_c^m)(B_{0z}+b_z) d\rho' dz'+\int_{0}^{R}\rho' D_c^m|_{z'=H}(B_{0\rho}+b_\rho) \nonumber\\
&&-\rho' D_s^m|_{z'=H}(B_{0\varphi}+b_\varphi)d\rho'-\int_{-H}^HR(B_{0z}+b_z)D_c^m|_{\rho'=R}dz'\nonumber\\
&&+\int_{0}^{R}\rho' D_s^m|_{z'=-H}(B_{0\varphi}+b_\varphi)-\rho' D_c^m|_{z'=-H} (B_{0\rho}+b_\rho)
d\rho']\label{eq12}\\
A_z&=& \frac{1}{4\pi}[\int_{-H}^H\int_{0}^{R}-\rho'\rho E_s^m (B_{0\rho}+b_\rho)+\rho'(\rho'E_1^m-\rho E_c^m)(B_{0\varphi}+b_\varphi)d\rho'dz'\nonumber\\
&&+\int_{-H}^H R D_1^m|_{\rho'=R}(B_{0\varphi}+b_\varphi) dz'],\label{eq13}
\end{eqnarray}
where the following abbreviations of azimuthal integrals were used:
\begin{eqnarray}
D_s^m(\rho,\rho',z,z')=\int_0^{2\pi}\frac{\sin \varphi' \sin m\varphi'}{(\rho^2-2\rho\rho' \cos \varphi'+\rho'^2+(z-z')^2)^{\frac{1}{2}}}d\varphi'\nonumber\\
D_c^m(\rho,\rho',z,z')=\int_0^{2\pi}\frac{\cos \varphi' \cos m\varphi'}{(\rho^2-2\rho\rho'\cos \varphi'+\rho'^2+(z-z')^2)^{\frac{1}{2}}}d\varphi'\nonumber\\
D_1^m(\rho,\rho',z,z')=\int_0^{2\pi}\frac{\cos m\varphi' }{(\rho^2-2\rho\rho'\cos \varphi'+\rho'^2+(z-z')^2)^{\frac{1}{2}}}d\varphi'.\nonumber
\end{eqnarray}

In our numerical scheme, we typically use equidistant grid points
$\rho_i=i\times \Delta r$ and $z_j=j\times \Delta z$  
to discretize the intervals $[0,R]$ and $[-H,H]$, respectively (in some
applications non-equidistant grid points are also used).
The extended trapezoidal rule is 
applied to approximate all the integrals in 
Eqs. (\ref{eq5}-\ref{eq13}). Then we obtain the following matrix equations
\begin{eqnarray}
\pmatrix{{{b}}_\rho\cr 
{{b}}_\varphi\cr
{{b}}_z\cr}&=&\mu\sigma \left[  {\mathbf{P}}\pmatrix{  {{B}}_{0\rho}+{{b}}_\rho   \cr
{{B}}_{0\varphi}+{{b}}_\varphi              \cr
{{B}}_{0z}+b_z  }-{\mathbf{Q}}\pmatrix{{{\phi}}_{s1}\cr
{{\phi}}_{s2}\cr
{{\phi}}_{s3}}-\lambda{\mathbf{R}}\pmatrix{{{A}}_\rho\cr 
{{A}}_\varphi\cr
{{A}}_z\cr} \right],\label{eq14a}\\
\frac{1}{2}\pmatrix{{\mathbf{\phi}}_{s1}\cr
{\mathbf{\phi}}_{s2}\cr
{\mathbf{\phi}}_{s3}}&=&{\mathbf{S}}\pmatrix{
{{B}}_{0\rho}+{{b}}_\rho   \cr 
{{B}}_{0\varphi}+{{b}}_\varphi              \cr
{{B}}_{0z}+b_z
}-\lambda{\mathbf{T}}\pmatrix{{{A}}_\rho\cr 
{{A}}_\varphi\cr
{{A}}_z}-{\mathbf{U}}\pmatrix{{\mathbf{\phi}}_{s1}\cr
{{\phi}}_{s2}\cr
{{\phi}}_{s3}},\label{eq14b}\\
\pmatrix{{{A}}_\rho\cr 
{{A}}_\varphi\cr
{{A}}_z}&=&{\mathbf{W}}\pmatrix{
{{B}}_{0\rho}+{{b}}_\rho   \cr 
{{B}}_{0\varphi}+{{b}}_\varphi              \cr
{{B}}_{0z}+b_z
   \cr},\label{eq14c}
\end{eqnarray}
where the matrix elements of $\mathbf{P}$, $\mathbf{Q}$, $\mathbf{R}$, $\mathbf{S}$, $\mathbf{T}$,
$\mathbf{U}$, and $\mathbf{W}$ can be read off from Eqs.(\ref{eq5}-\ref{eq13}).
Combining Eqs.(\ref{eq14a}-\ref{eq14c}), we obtain
\begin{eqnarray}{\label{eq16a}}
({\mathbf{I}}-\mu\sigma {\mathbf{E}}-\mu\sigma\lambda{\mathbf{F}}){\mathbf{b}}=\mu\sigma({\mathbf{E}}+\lambda{\mathbf{F}}){\mathbf{B}}_0,
\end{eqnarray}
where
\begin{eqnarray}
{\mathbf{E}}&=&{\mathbf{P}}-{\mathbf{Q}} \cdot(\frac{1}{2}{\mathbf{I}}+{\mathbf{U}})^{-1} \cdot {\mathbf{S}}  \label{eq16b}\\
{\mathbf{F}}&=&{\mathbf{Q}} \cdot (\frac{1}{2}{\mathbf{I}} + {\mathbf{U}})^{-1} \cdot {\mathbf{T}} \cdot {\mathbf{W}} - {\mathbf{R}} \cdot {\mathbf{W}} \; .\label{eq16c}
\end{eqnarray}
After solving the algebraic equation system (\ref{eq16a}), the
induced magnetic field $\textbf{b}$ can be obtained for the magnetic induction process. 

For the kinematic dynamo problem, the following generalized eigenvalue problem has to be solved
\begin{eqnarray}{\label{eq17}}
({\mathbf{I}}-\mu \sigma {\mathbf{E}})\cdot{\mathbf{b}}=\lambda^*{\mathbf{F}} \cdot {\mathbf{b}}
\end{eqnarray}
for the given velocity field $\textbf{u}$, where $\lambda^*=\mu\sigma\lambda$. 
Note that a quite similar numerical scheme can be established  in
spherical geometry for the case of  axisymmetric dynamo sources.

\section{NUMERICAL IMPLEMENTATION AND RESULTS}
In this section, the integral equation approach  will be applied to 
various cylindrical dynamo and induction problems of experimental relevance.
We start with the problem of
the  free decay of a magnetic field in a cylinder. Then,
a class of Beltrami-like flows will be considered.
In all the problems we use the QZ algorithm \cite{MOLER} which is 
a modification  of the QR algorithm for the case of generalized non-hermitian 
eigenvalue problems. 

The integral equation approach is further employed to investigate the 
induction effect of the VKS experiment. The algebraic equation system 
is solved by the LU
decomposition.
The obtained induced magnetic field will be compared with 
the data measured in experiment.

At the end we deal with the Riga dynamo experiment with its large ratio
of height to radius. Due to the large resulting matrices 
we shift here from direct matrix inversion methods to
the generalized inverse iteration method \cite{IVERS}.

\subsection{Free field decay in a finite cylinder}
\begin{figure}
{\includegraphics[width=12cm]{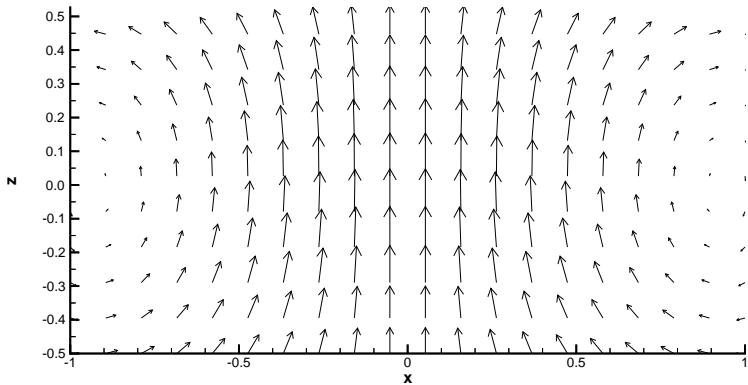}}
\caption{Freely decaying magnetic field in a finite cylinder with $R=2H=1.$}
\label{fd}
\end{figure}

The simplest problem to start with is the free decay of a magnetic
field in a finite length cylinder. This example was already treated
by Iskakov et al. \cite{ISKAKOV1}. In Fig. \ref{fd} we show the
magnetic field
lines of the slowest decaying eigenfield, which has the same 
dipolar structure as
in Fig. 8 in \cite{ISKAKOV1}.

\subsection{Beltrami-like flows}
In this subsection, we consider a class of flows in finite cylinders
which we call ''Beltrami-like'' flows. Actually, the velocity field 
${\mathbf v}({\mathbf r})$ of Beltrami flows are characterized by the property 
$\nabla \times {\mathbf v}=\beta {\mathbf v}$. In dynamo theory, there is
particular interest in such flows since they are
also helicity maximizing.
Helicity maximizing flows, in turn, are well known to possess
quite small critical $R_m$, a fact that was utilized, e.g.,
in the optimization of the
Riga dynamo experiment \cite{Stef1}. 
The actual flow structures that will be treated in this work were
proposed by 
J. L\'{e}orat \cite{LEORAT}, and a certain sub-class of them (with 
ideally conducting boundary conditions, however) 
was considered  by Wang et al. \cite{WANG}.

\begin{figure}
{\includegraphics[width=14cm]{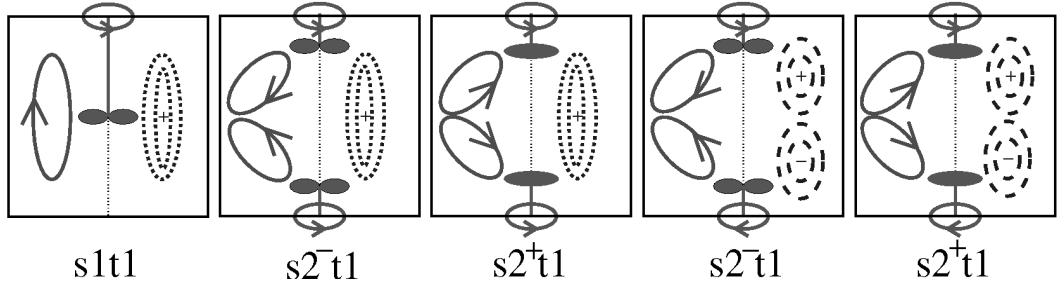}}
\caption{Illustration of the considered flow topologies $s_m^{\pm}t_n$, and
of typical propeller or rotating disk configurations to produce them.}
\label{topo}
\end{figure}

We use the notation $s_m^{\pm}t_n$ to characterize flows
with $m$
poloidal vortices and $n$ toroidal vortices. The
sign $\pm$ indicates that the poloidal flow in the
equatorial plane is directed inward ($+$) or outward ($-$), respectively.
An impression of the topology of the flow structure can be
obtained from Fig. \ref{topo} where we have also indicated
possible propeller or rotating disk configurations to produce such flows.

The analytical expression of the flows to be considered in this paper
are as follows:
\begin{eqnarray}{\label{bl1}}
v_r(r,z)&=&c_1 \; J_1(\alpha \; r) \cos(m\pi (z+H)/2H),\\
v_\varphi(r,z)&=&\tau \; J_1(\alpha \; r) \cos(n\pi(z+H)/2H),\\
v_z(r,z)&=&-c_1 \; c_2 \; \alpha/\pi \; J_0(\alpha \; r) \sin(m\pi(z+H)/2H),
\end{eqnarray}
where $\alpha=3.8317$ is the first root of the Bessel function $J_1$. 
In the following, we will restrict ourselves to $m,n=1,2$. $c_1=1$
for all $s_m^+t_n$ flows, $c_1=-1$ for all $s_m^-t_n$ flows,
$c_2=H/2$ for $s_1t_1$ flow and for other flows $c_2=1$.  Again, $2 H$ is
the height of the cylinder, $z\in[-H,H]$. In the following discussion
the height of the cylinder is set to $2$ and
the radius is fixed to $1$.
The parameter $\tau$ indicates the ratio of
toroidal to poloidal flow. We will consider values of $\tau$ close to
2 which turned out to be advantageous for dynamo action. Actually,
this value $\tau=2$ is not the value which would
correspond to an excact Beltrami flow.
This is the reason why we have called the
considered flows ''Beltrami-like''.

In what follows we will use a definition of the magnetic Reynolds
number $R_m$ which is based on the maximum of the axial velocity. 
In order to display  the results for the  $s_m^{\pm}t_n$ in one
common figure we will use the ${\pm}$ as a sign of $R_m$ according to
$R_m=\pm \mu \sigma R |v^{max}_z|$.

For the case without external layer ($w=0$) we found that
only the $s_2^+t_2$ dynamo is steady,
all the others are all oscillatory. However, if an external
layer around the finite cylinder is added, even the thickness of the
layer is quite small, for example, equal to $0.05$, the $s2^-t2$ dynamo 
becomes steady.

\begin{figure}
\begin{center}
\begin{tabular}{c}
\includegraphics[width=0.9\textwidth]{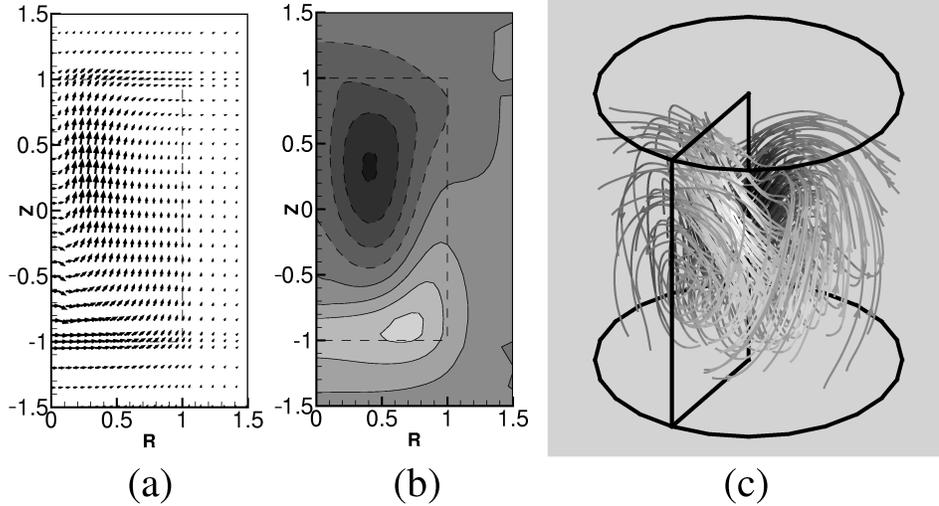}
\end{tabular}
\end{center}
\caption{Magnetic field structure for $s_1t_1$ flow with $w=0.5$. (a)
Poloidal field component. (b) Contour
plot of the toroidal field component. (c) Three dimensional field structure.}
\label{s1t1f}
\end{figure}

\begin{figure}
\begin{center}
\includegraphics[width=0.9\textwidth]{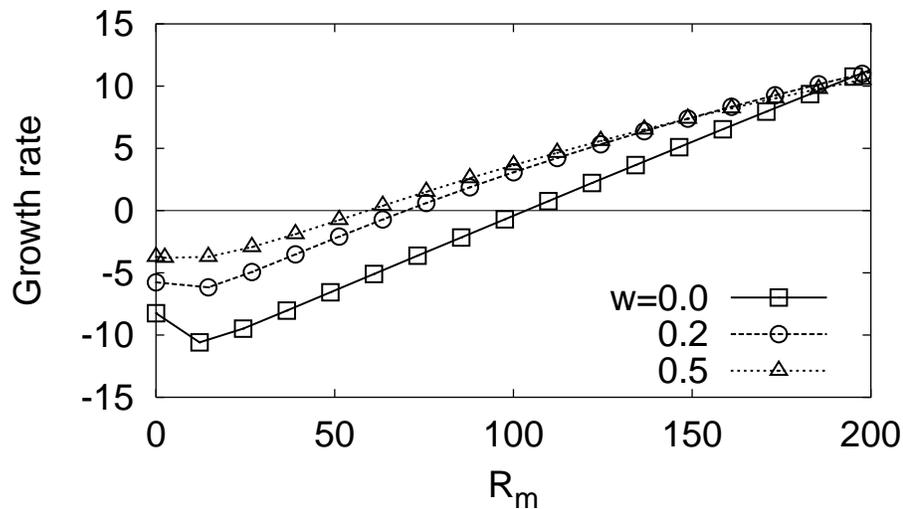}
\caption{Growth rates for $s_1t_1$ flow and the influence of the externally added layers
with thickness $w$.}
\label{s1t1g}
\end{center}
\end{figure}

\begin{figure}
\begin{center}
\begin{tabular}{c}
\includegraphics[width=0.9\textwidth]{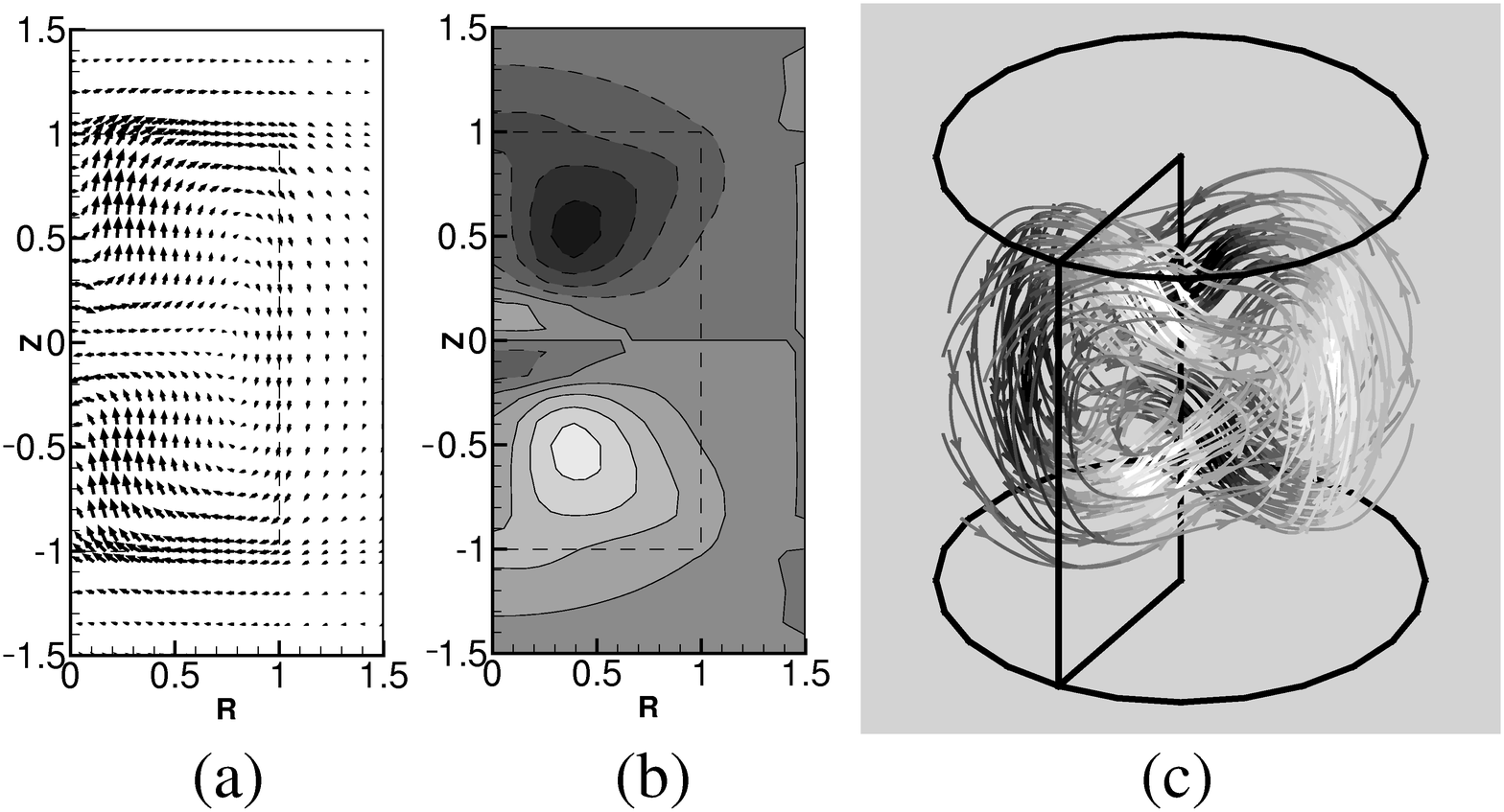}
\end{tabular}
\end{center}
\caption{Magnetic field structure for $s_2^-t_1$ flow}
\label{s2mt1f}
\end{figure}

\begin{figure}
\begin{center}
\begin{tabular}{c}
\includegraphics[width=0.9\textwidth]{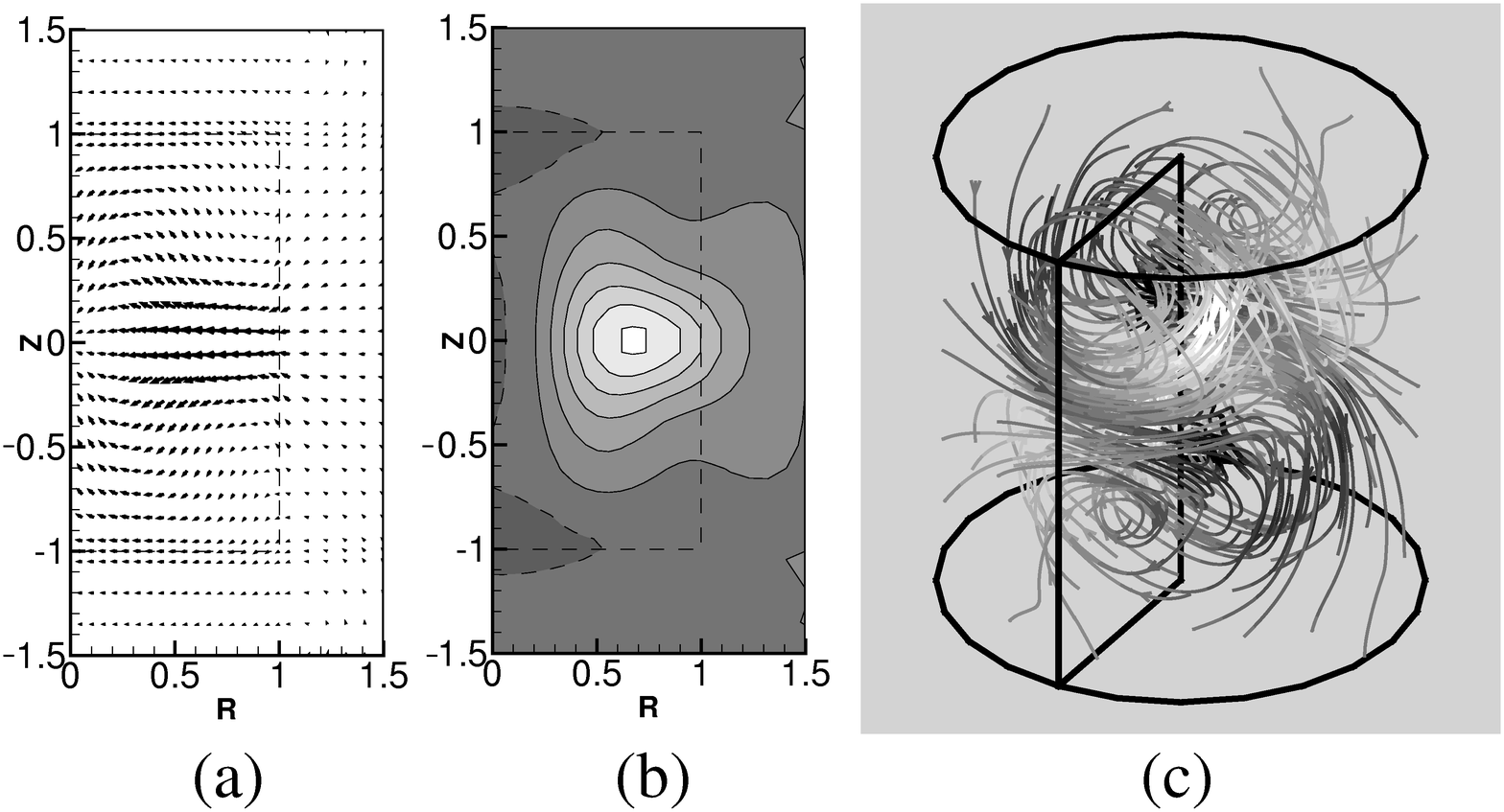}
\end{tabular}
\end{center}
\caption{Magnetic field structure for $s_2^+t_1$ flow.}
\label{s2pt1f}
\end{figure}

\begin{figure}
\begin{center}
\includegraphics[width=0.9\textwidth]{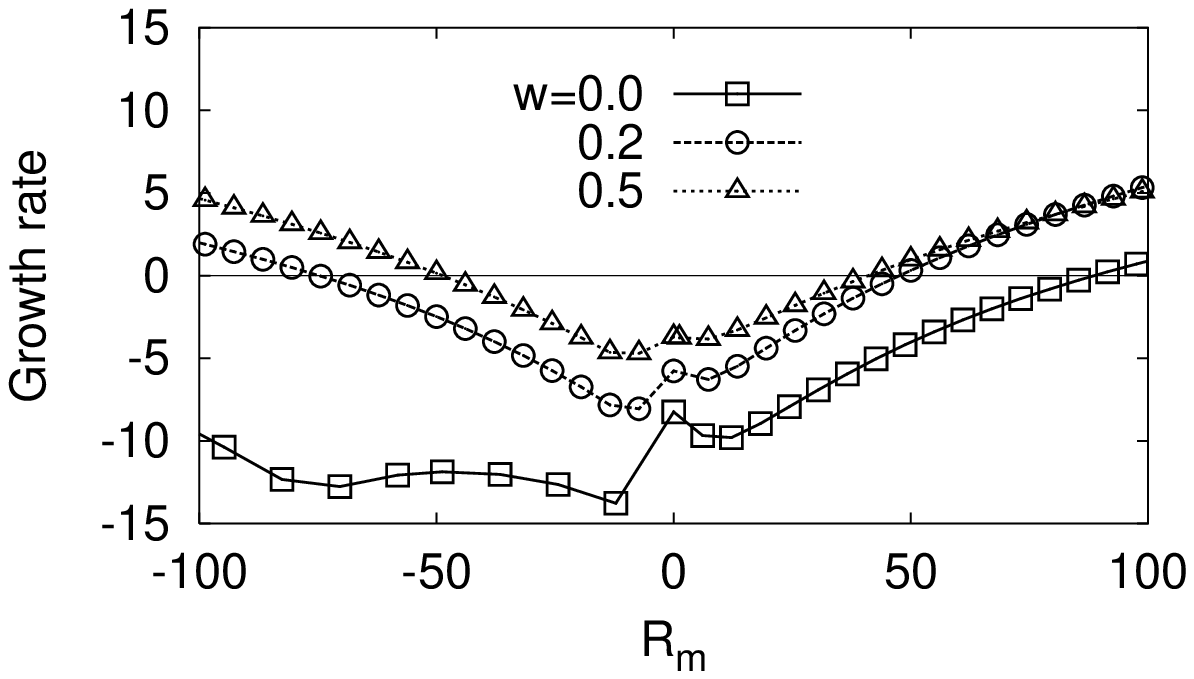}
\caption{Growth rates for $s_2^{\pm}t_1$ flow and influence of the externally added layers.}
\label{s2t1g}
\end{center}
\end{figure}

\begin{figure}
\begin{center}
\begin{tabular}{c}
\includegraphics[width=0.9\textwidth]{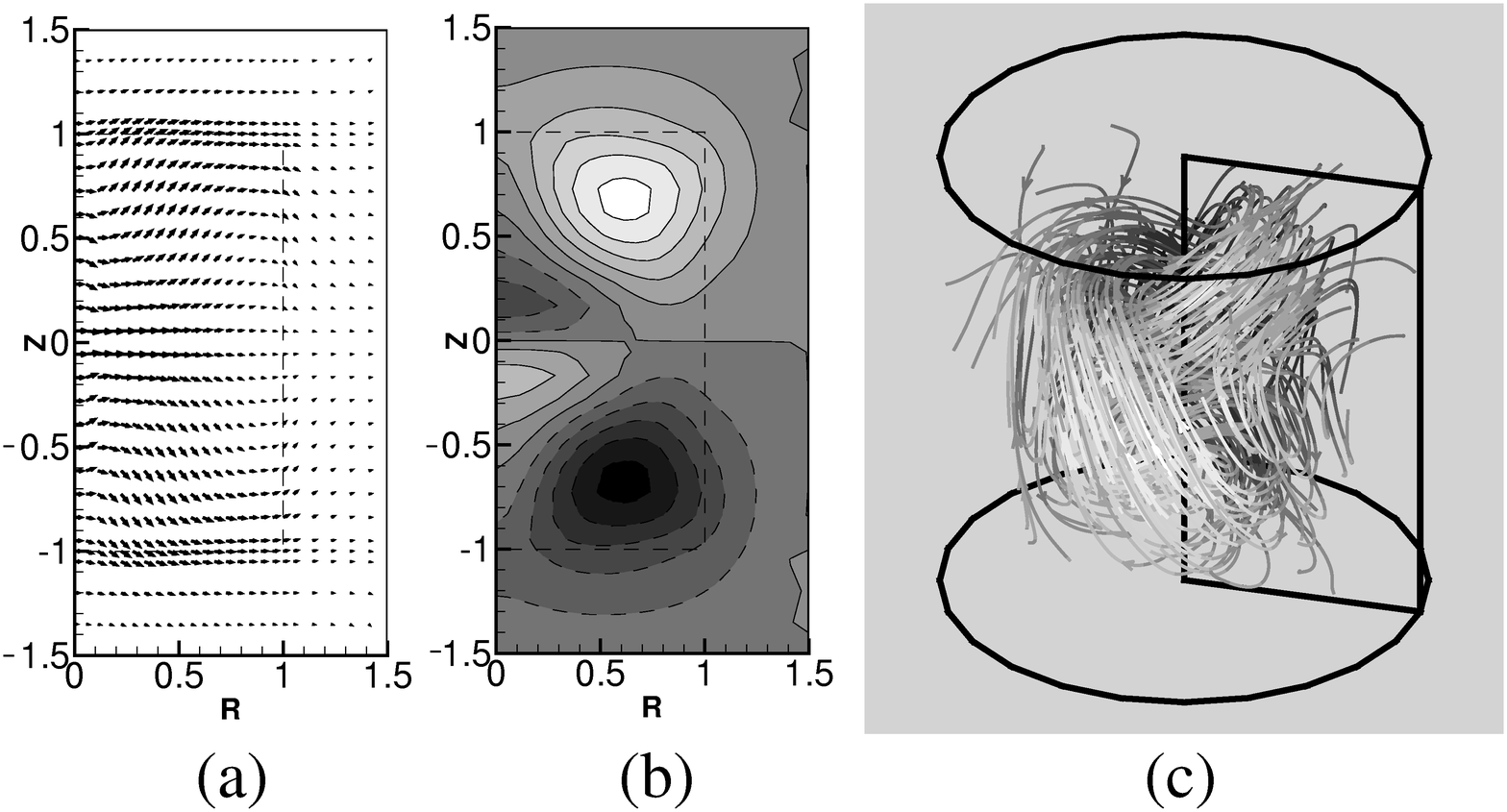}
\end{tabular}
\end{center}
\caption{Magnetic field structure for $s_2^-t_2$ flow}
\label{s2mt2f}
\end{figure}

\begin{figure}
\begin{center}
\begin{tabular}{c}
\includegraphics[width=0.9\textwidth]{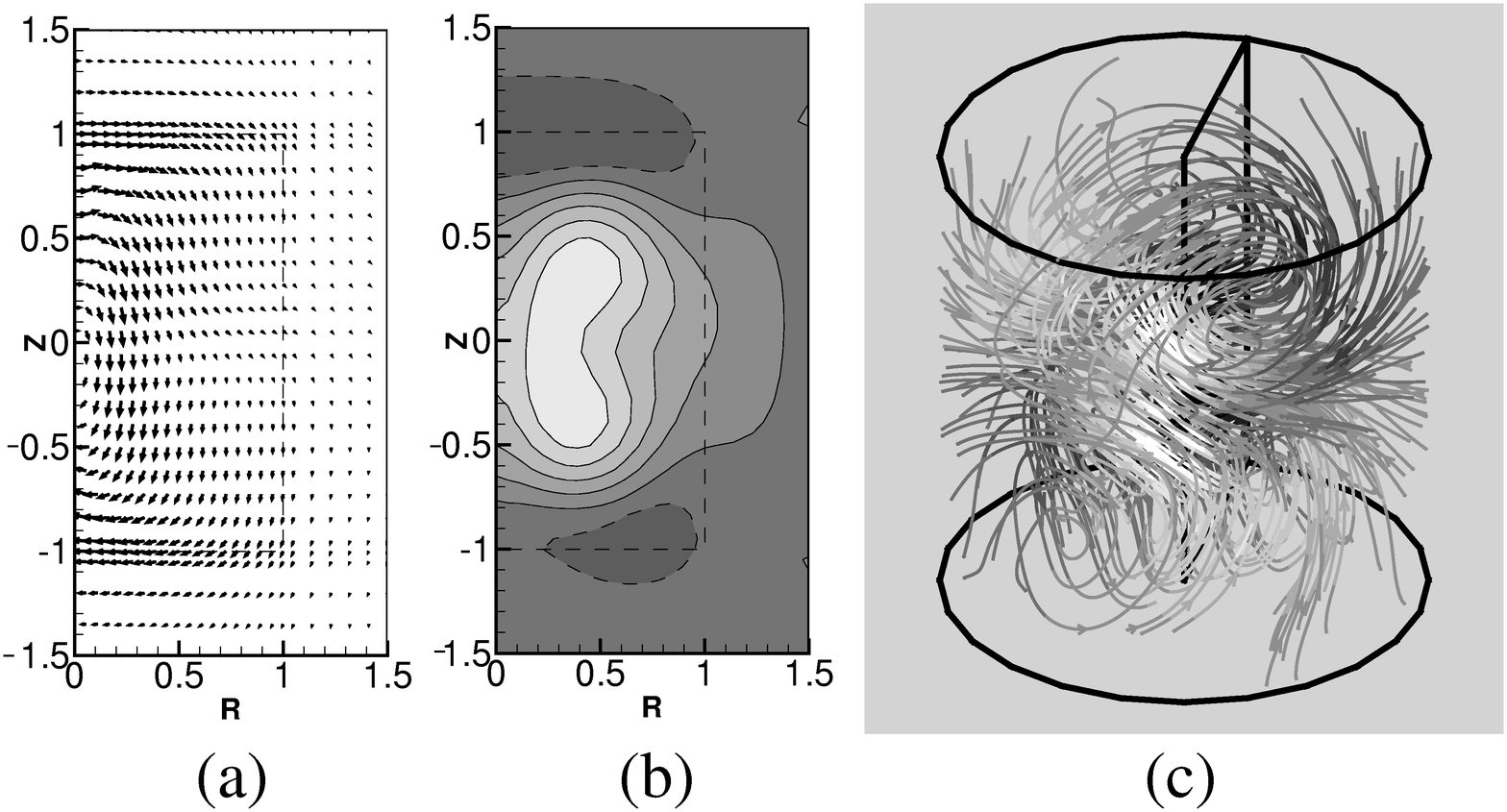}
\end{tabular}
\end{center}
\caption{Magnetic field structure for $s_2^+t_2$ flow.}
\label{s2pt2f}
\end{figure}

\begin{figure}
\begin{center}
\includegraphics[width=0.9\textwidth]{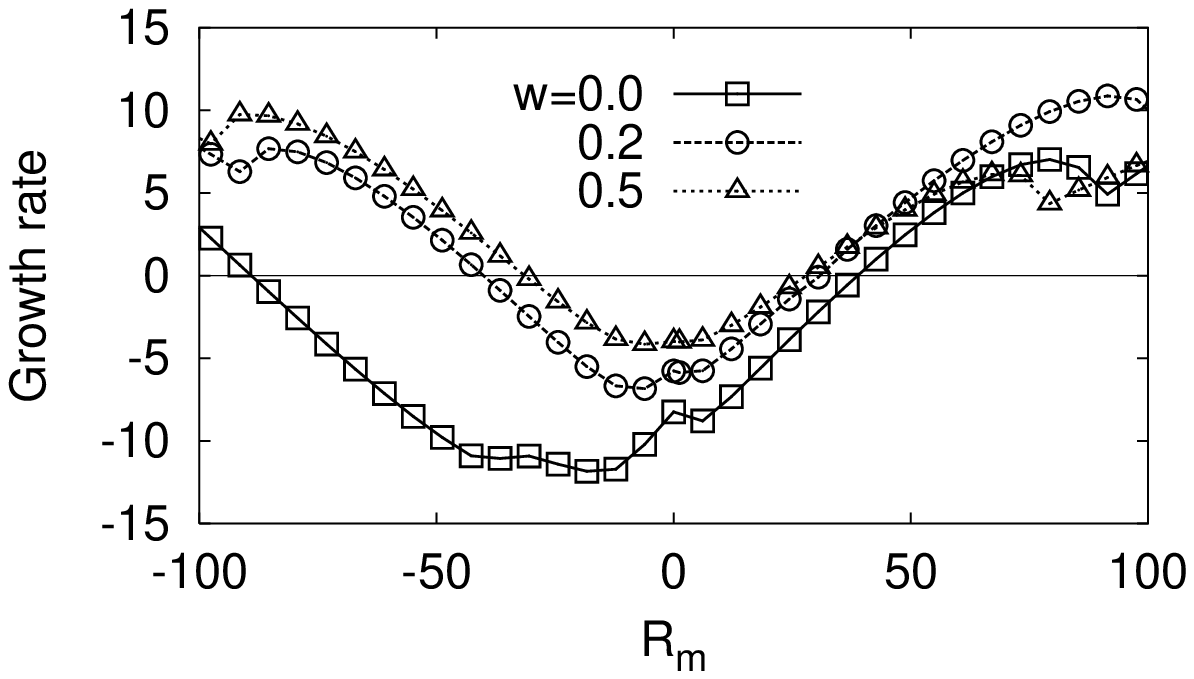}
\caption{Growth rates for $s_2^{\pm}t_2$ flow and influence of the externally added layers}
\label{s2t2g}
\end{center}
\end{figure}

The magnetic field structures for $s_1t_1$, $s_2^{-}t_1$, $s_2^{+}t_1$, $s_2^{-}t_2$ and
$s_2^{+}t_2$ flows are shown, at an azimuthal section at $\varphi=0$, in
Figs. \ref{s1t1f}, \ref{s2mt1f}, \ref{s2pt1f}, \ref{s2mt2f} and \ref{s2pt2f},
respectively. In all these cases, an externally added layer with 
thickness equal to $0.5$ has been considered. The variations of growth 
rates of the magnetic fields with respect to the magnetic Reynolds number 
for all the flows are depicted in Figs. \ref{s1t1g}, \ref{s2t1g} and \ref{s2t2g}.
From these figures, one can see that the externally added layer has a very strong 
impact on the onset of dynamo actions. For example, the critical magnetic Reynolds 
number for the flow $s_2^{-}t_1$ is approximately equal to
$143$ in the
case without external layer. When an external layer with thickness $0.2$ is 
considered, the critical magnetic Reynolds number reduces to $61$. If the 
thickness of the external layer is increased to $0.5$, the critical magnetic 
Reynolds number further declines to $40$. Finally, one can also note that 
there is a tendency that when the thickness of the external layer becomes 
larger, the curves of the growth rates  become more symmetric with respect 
to the
ordinate axis at $R_m=0$.

%\foilhead{Results: S2+T2, Magnetic field structure}
%\foilhead{Results: s2$^{\pm}$t2, Growth rates for various layers}

\begin{figure}
\begin{center}
\includegraphics[width=0.9\textwidth]{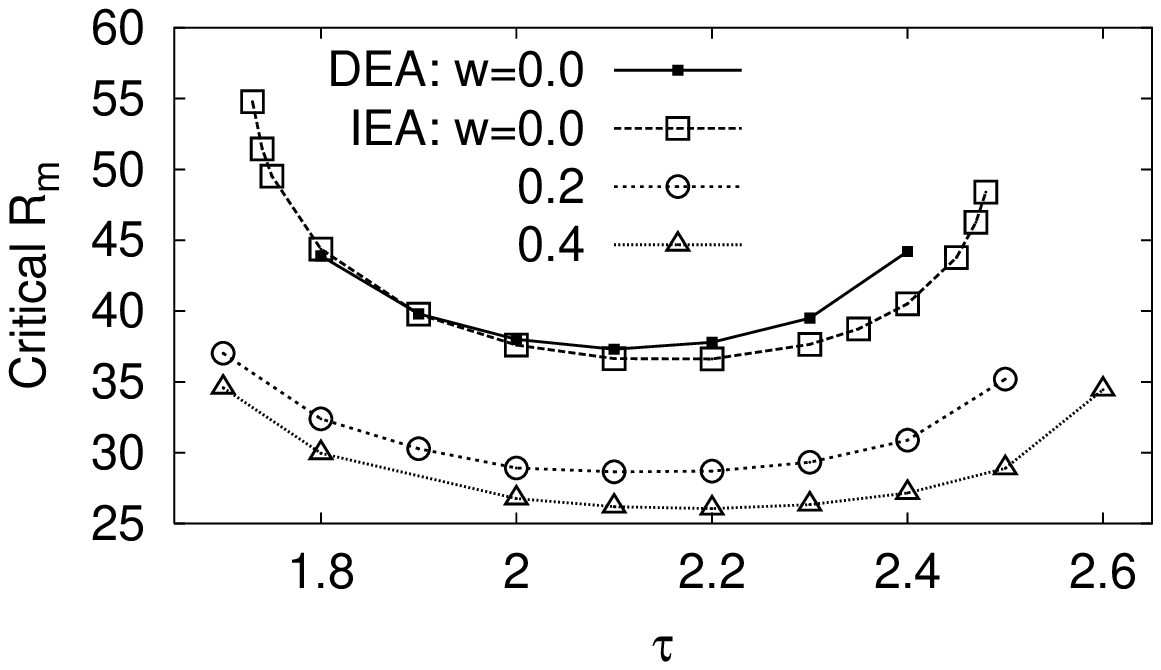}
\caption{Critical $R_m$ for the $s_2^+t_2$ flow in dependence on the
toroidal/poloidal ratio $\tau$ for different thicknesses $w$ of the surrounding
layer. For $w=0$, we compare the results with the results of a
differential equation solver.}
\label{torpol}
\end{center}
\end{figure}

In Fig. \ref{torpol} we show, for the flow $s_2^+t_2$ with different
widths $w$ of the external
layer, 
the dependence of the critical $R_m$ on the
parameter $\tau$ which measures the ratio of toroidal to poloidal motion.
For $w=0$ we show, in addition to the results of the integral equation approach,
also the results of a finite difference code based on the differential
equation approach as it was described in \cite{Stef1} and also used in \cite{Stef3}. 
In general, we observe a good correspondence of the
results of both
methods which, however, deteriorates slightly for increasing values of $\tau$.

\subsection{Induction effects in the VKS Experiment}
Since the $s_2^+t_2$ flows are characterized by a comparatively low
critical $R_m$, much focus was laid on their realization 
in experimental dynamos. Both the spherical  Madison
dynamo experiment (MDX) \cite{MDX1,MDX2} and the cylindrical
''von K\'{a}rm\'{a}n sodium'' experiment (VKS) \cite{VKS1,VKS2}
are realizations
of the $s_2^+t_2$ flow.

Although a recent version (using impellers with a high
magnetic permeability $\mu_{rel}\sim 200$) of the VKS
experiment has shown dynamo action \cite{Monc} and even
a kind of polarity reversals \cite{Berh}, the under-performance
of the original VKS experiment compared to numerical predictions
is still a matter of interest. In \cite{Stef3} we have tried
to explain this by the detrimental effect of sodium layers
behind the propellers on the dynamo action. The sheer existence
of these layers leads already to a significant increase of the
critical $R_m$ which becomes dramatic if a
realistic
rotation therein is taken into account.

Interestingly, in the original VKS experiment,
the measured induced magnetic
fields, for large $R_m$,
are significantly weaker than the numerically predicted ones.
Using our method we will try to figure out if this effect can also be
attributed to the existence of lid layers and the flow therein.
\begin{figure}
\begin{tabular}{c}
{\includegraphics[width=0.9\textwidth]{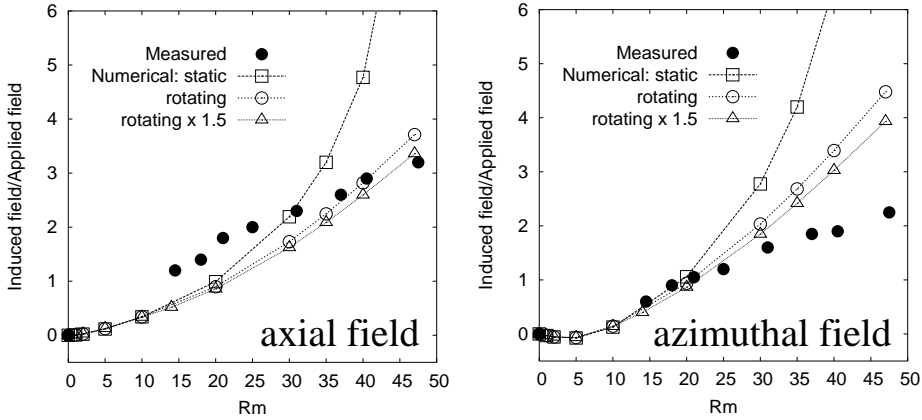}}
\end{tabular}
\caption{Ratio of induced axial (left) and azimuthal (right) 
magnetic field to the applied magnetic
field. 
Experimental results at radius $r=0.5$ (taken from \cite{RAV2})
and numerical results under the assumptions of different flows
in the lid layer.
Static lid layer, constant velocity in lid layer, constant
velocity in lid layers
multiplied by factor 1.5.}
\label{vksaxial}
\end{figure}

A rather realistic flow field, resulting from the so-called TM73
propeller\cite{Rave}  (which
was identified as
a sort of optimal
flow field),
is considered in our calculation.
Some interpolations were necessary
to project this flow field onto the grids used 
in our code. More details on this can be found in \cite{Stef3}.
\begin{figure}
\begin{tabular}{c}
{\includegraphics[width=12cm]{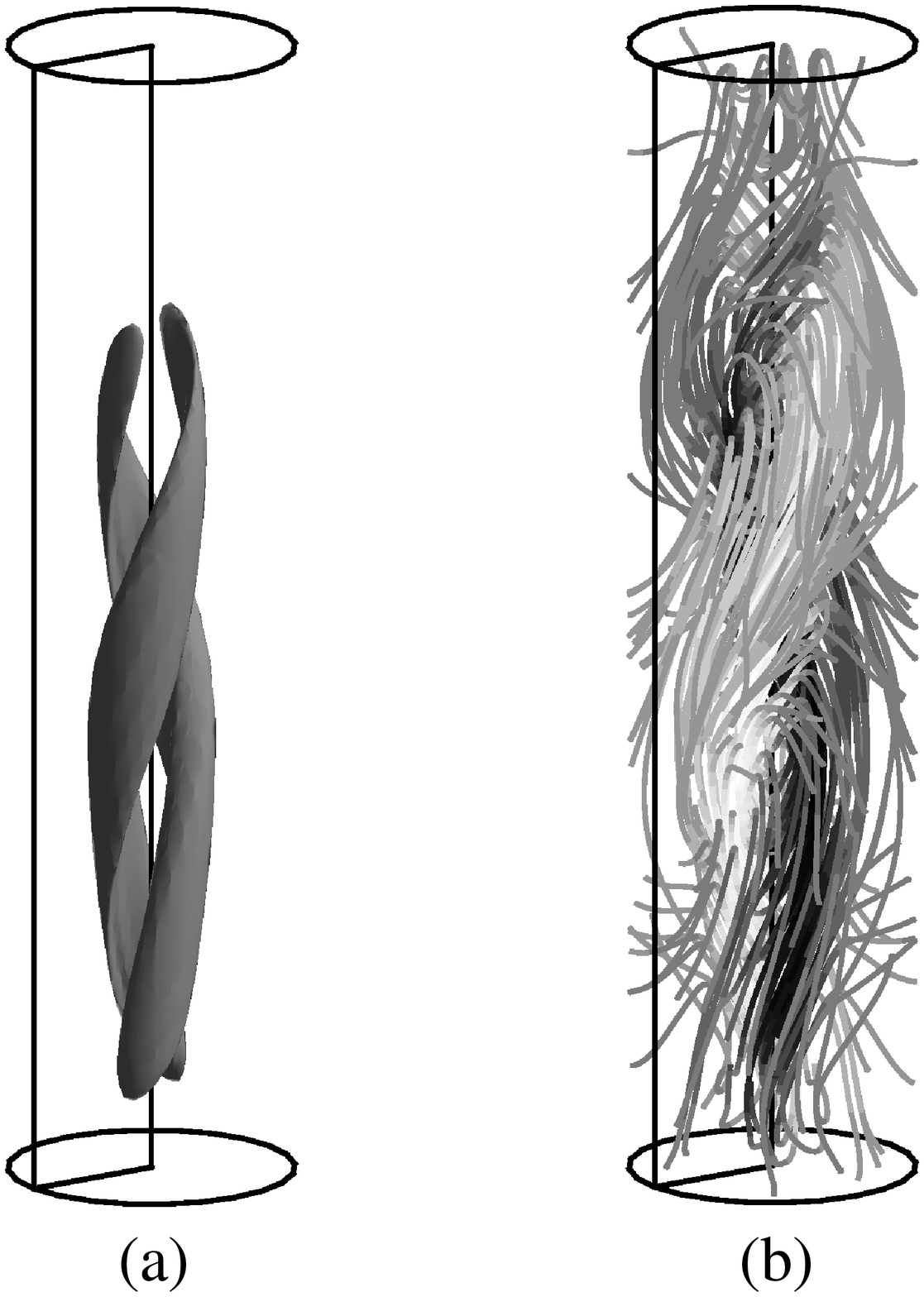}}
\end{tabular}
\caption{Simulated magnetic field structure of the eigenmode of the Riga dynamo 
experiment. (a) Isosurface plot of the magnetic field energy. The 
isosurface corresponds to 25 per cent of the maximum magnetic energy.
(b) Magnetic field lines.}
\label{rigaf}
\end{figure}

In the original VKS experiment a static external magnetic field
was applied in the transverse direction.
The induced magnetic field was measured in the direction perpendicular
to the externally
added magnetic field at the point $r=0.5$ in the equatorial
plane. In the following the
induced magnetic fields near the $r=0.5$ points obtained by
our integral equation approach
are compared with the measured ones. The influence
of a rotating flow in the lid
layer on the induced field is
investigated. Three kinds of velocity field in the lid layer
are considered. The first one
is 
a static lid layer. The second one is that only a
rotation of the lid layer is assumed,
but $v_\varphi$ remains constant in the axial direction, its dependence on the radial
variable is the same as on the interface between the lid layer and inner part of the cylinder. 
The third one has only one difference from the second
case in that the magnitude of $v_\varphi$
is increased by a factor $1.5$ which comes closer to the velocity of the impeller.
For these three cases, the numerical axial induced fields
around the point $r=0.5$ and the experimental result at $r=0.5$ on the
equatorial plane are
shown in the left part of Fig. \ref{vksaxial} for
different magnetic Reynolds numbers. The right panel of Fig.
\ref{vksaxial} displays the azimuthal
induced magnetic fields. From these figures one
can see that the third case shows a best
agreement with the experimental one. A good agreement of the axial magnetic field with the 
experimental result has been achieved for the second and third cases. But for the azimuthal 
magnetic field, when the magnetic Reynolds number is larger than $30$, there is still a gap 
between the numerical results and the experimental ones.

Nevertheless it is quite likely that it is indeed the existence of lid layers
and some azimuthal flow therein which is responsible for the
unexpected under-performance of the original VKS dynamo experiment.
\subsection{Riga experiment}
In this subsection, the integral equation approach is used to re-simulate the 
kinematic regime of the Riga dynamo experiment. This experiment has been 
optimized and analyzed extensively within the differential equation 
approach (DEA) by means of
a finite difference solver \cite{Stef1}.
The values of the velocity field on the grids used in our 
code are obtained by interpolating the experimental velocity field 
measured in 
a water-dummy experiment. The influence of less conducting 
stainless steel walls has not been taken into account. 

The computations have been carried out on  a 100$\times$20 grid in z- and r-direction. 
The structure of the magnetic eigenfield is illustrated 
in Fig. \ref{rigaf}. Figure \ref{rigaf}a shows the isosurface of the
magnetic field energy  (this time at 25  percent of the
maximum value). 
In Fig. \ref{rigaf}b the magnetic eigenfield lines 
are depicted. Basically the structure is the same as that resulting from the differential 
equation approach  \cite{PLASMA}  with a 401$\times$64 grid in z- and r-direction.
\begin{figure}
\begin{tabular}{ccc}
{\includegraphics[width=14cm]{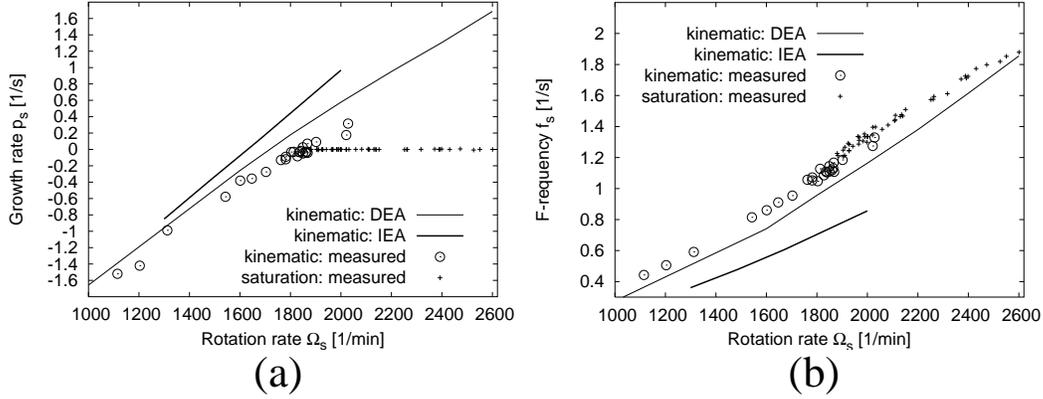}}
\end{tabular}
\caption{Comparison of the IEA and DEA results for the Riga dynamo experiment,
together with experimental results. (a) Growth rate. (b) Frequency.}
\label{rigagr}
\end{figure}

The dependence of the growth rate 
and frequency of the eigenmode of the 
Riga dynamo experiment on the rotation rate is 
shown in Fig. \ref{rigagr}a and Fig.\ref{rigagr}b, respectively. The comparison with the DEA results 
shows that the slopes of the curves are in good agreement. 
However, we see that the limited grid resolution in the IEA leads 
to significant shifts in the order of 5 percent towards lower rotation rates 
for the growth rate 
and of 10 per cent towards higher rotation rate for the frequency.
Hence, it could be said  that the Riga dynamo experiment marks 
a margin of reasonable   applicability of 
the IEA  with  its need to invert large matrices which 
are fully occupied.
\section{CONCLUDING REMARKS}
In the present paper, the integral equation approach to   
kinematic dynamos has been applied to non-spherical geometries. 
The method was
examined by its application to the free field decay. The comparison 
of the obtained results with other methods shows a good agreement. 
The integral equation approach was extended 
to investigate induction effects of the VKS experiment. 
The obtained induced magnetic field shows a satisfactory 
agreement with the experimental result when the effect of the lid layers and
a certain azimuthal flow therein are taken into account. 
Finally, it was   applied to simulate the Riga dynamo experiment.

It can be concluded that the 
integral equation approach is robust and reliable and can be 
used for practical purposes, although limits of its
applicability are seen for the Riga dynamo experiment with
its large ratio of length to radius.

\end{document}